\documentclass[11pt, a4paper]{article}

\usepackage{url}
\usepackage{graphicx}
\usepackage{textcomp}
\usepackage{authblk}
\usepackage{amssymb}
\usepackage{amsthm}
\usepackage{ulem}
\usepackage{tabularx}

\begin{document}
\title{Cosmic Ray Measurements with LOPES: Status and Recent Results}


\author[1]{F.G.~Schr\"oder\textsuperscript{\footnotesize{*}}}
\author[1]{W.D.~Apel}
\author[2]{J.C.~Arteaga}
\author[3]{L.~B\"ahren}
\author[1]{K.~Bekk}
\author[4]{M.~Bertaina}
\author[5]{P.L.~Biermann}
\author[1,6]{J.~Bl\"umer}
\author[1]{H.~Bozdog}
\author[7]{I.M.~Brancus}
\author[4]{A.~Chiavassa}
\author[1]{K.~Daumiller}
\author[9]{V.~de~Souza}
\author[4]{F.~Di~Pierro}
\author[1]{P.~Doll}
\author[1]{R.~Engel}
\author[3,10,5]{H.~Falcke}
\author[6]{B.~Fuchs}
\author[11]{D.~Fuhrmann}
\author[12]{H.~Gemmeke}
\author[13]{C.~Grupen}
\author[1]{A.~Haungs}
\author[1]{D.~Heck}
\author[3]{J.R.~H\"orandel}
\author[5]{A.~Horneffer}
\author[6]{D.~Huber}
\author[1]{T.~Huege}
\author[14]{P.G.~Isar}
\author[11]{K.-H.~Kampert}
\author[6]{D.~Kang}
\author[12]{O.~Kr\"omer}
\author[3]{J.~Kuijpers}
\author[6]{K.~Link}
\author[15]{P.~{\L}uczak}
\author[6]{M.~Ludwig}
\author[1]{H.J.~Mathes}
\author[6]{M.~Melissas}
\author[8]{C.~Morello}
\author[1]{J.~Oehlschl\"ager}
\author[6]{N.~Palmieri}
\author[1]{T.~Pierog}
\author[11]{J.~Rautenberg}
\author[1]{H.~Rebel}
\author[1]{M.~Roth}
\author[12]{C.~R\"uhle}
\author[7]{A.~Saftoiu}
\author[1]{H.~Schieler}
\author[12]{A.~Schmidt}
\author[16]{O.~Sima}
\author[7]{G.~Toma}
\author[8]{G.C.~Trinchero}
\author[1]{A.~Weindl}
\author[1]{J.~Wochele}
\author[15]{J.~Zabierowski}
\author[5]{J.A.~Zensus}

\affil[1]{\small{Institut f\"ur Kernphysik, Karlsruhe Institute of Technology (KIT), Germany}}
\affil[{2}]{Universidad Michoacana, Morelia, Mexico}
\affil[3]{Radboud University Nijmegen, Department of Astrophysics, The Netherlands}
\affil[4]{Dipartimento di Fisica Generale dell' Universit\`a Torino, Italy}
\affil[5]{Max-Planck-Institut f\"ur Radioastronomie Bonn, Germany}
\affil[6]{Institut f\"ur Experimentelle Kernphysik, Karlsruhe Institute of Technology (KIT), Germany}
\affil[7]{National Institute of Physics and Nuclear Engineering, Bucharest, Romania}
\affil[8]{INAF Torino, Instituto di Fisica dello Spazio Interplanetario, Italy}
\affil[{9}]{Universidad S\~ao Paulo, Inst. de F\'{\i}sica de S\~ao Carlos, Brasil}
\affil[10]{ASTRON, Dwingeloo, The Netherlands}
\affil[{11}]{Universit\"at Wuppertal, Fachbereich Physik, Germany}
\affil[{12}]{Institut f\"ur Prozessdatenverarbeitung und Elektronik, Karlsruhe Institute of Technology (KIT), Germany}
\affil[13]{Universit\"at Siegen, Fachbereich Physik, Germany}
\affil[{14}]{Institute for Space Sciences, Bucharest, Romania}
\affil[{15}]{National Centre for Nuclear Research, Department of Cosmic Ray Physics, {\L}\'{o}d\'{z}, Poland}
\affil[{16}]{University of Bucharest, Department of Physics, Romania}

\date{\normalsize{The following article has been accepted by AIP Conference Proceedings.}\\Proceedings of ARENA 2012, Erlangen, Germany\\ \hspace{0.5 cm} \small{to appear on \url{http://proceedings.aip.org/}}}

\maketitle

\clearpage

\begin{abstract}
LOPES is a digital antenna array at the Karlsruhe Institute of Technology, Germany, for cosmic-ray air-shower measurements. Triggered by the co-located KASCADE-Grande air-shower array, LOPES detects the radio emission of air showers via digital radio interferometry. We summarize the status of LOPES and recent results. In particular, we present an update on the reconstruction of the primary-particle properties based on almost 500 events above $100\,$PeV. With LOPES, the arrival direction can be reconstructed with a precision of at least $0.65^\circ$, and the energy with a precision of at least $20\,\%$, which, however, does not include systematic uncertainties on the absolute energy scale. For many particle and astrophysics questions the reconstruction of the atmospheric depth of the shower maximum, $X_\mathrm{max}$, is important, since it yields information on the type of the primary particle and its interaction with the atmosphere. Recently, we found experimental evidence that the slope of the radio lateral distribution is indeed sensitive to the longitudinal development of the air shower, but unfortunately, the $X_\mathrm{max}$ precision at LOPES is limited by the high level of anthropogenic radio background. Nevertheless, the developed methods can be transferred to next generation experiments with lower background, which should provide an $X_\mathrm{max}$ precision competitive to other detection technologies.
\end{abstract}

\section{Introduction}
Radio emission from air showers is studied both experimentally and theoretically already for about 50 years, and in the last years significant progress has been achieved. Analyses with digital antenna arrays like LOPES could show that at least the energy and the arrival direction can be reconstructed precisely enough to make radio arrays a useful tool for cosmic-ray physics above $10^{17}\,$eV. These results are generally confirmed by other experiments, e.g., CODALEMA \cite{ArdouinBelletoileCharrier2005, LautridouARENA2012} and AERA \cite{Melissas_AERA_ARENA2012, GlaserARENA2012}. It still remains to be demonstrated that radio measurements can contribute significantly to disentangle the composition of ultra-high energy cosmic rays. Recently, we could show that LOPES measurements are sensitive to the longitudinal shower development, and thus -- at least on a statistical level -- to the composition of the primary cosmic rays. However, the currently achieved precision on the atmospheric depth of the shower maximum, $X_\mathrm{max}$, is insufficient due to the high level of anthropogenic radio background at the LOPES site.

Since the LOPES experiment with its different setups has been described in several previous publications (e.g., reference \cite{FalckeNature2005, HuegeARENA_LOPESSummary2010, ApelLOPES3D2012}), we will give only a short summary of our setup and the analysis procedures, here. LOPES is a digital antenna array co-located with and triggered by the KASCADE-Grande experiment \cite{Apel2010KASCADEGrande} at the Karlsruhe Institute of Technology in Germany. It was built in 2003 as LOFAR \cite{NellesARENA_LOFAR2012} prototype station and made the proof-of-principle that air showers can be measured with digital radio interferometry \cite{FalckeNature2005}.

Because of its success, LOPES was enhanced several times. In 2005, it was extended from 10 to 30 inverted v-shape dipole antennas, which were all aligned in the east-west direction, because we assumed that the radio signal is mainly east-west polarized. End of 2006, half of the antennas were rotated to north-south alignment, and thus we could test the previous assumption -- it is true at least for most shower geometries \cite{IsarArena2008}. In 2009, we started LOPES-3D \cite{ApelLOPES3D2012}, which consists of 10 tripole antennas (= three crossed dipoles in the same location), to study the polarization in more detail. 

\vspace{\itemsep}
Generally the analyses and developments performed at LOPES aim at three main goals:

\vspace{\itemsep} \noindent \textbf{Reconstruction of air shower parameters:}\\
Already in the 1960s historic experiments based on analog techniques demonstrated that in principle the arrival direction and the primary energy can be reconstructed from radio measurements \cite{Allan1971}, but they could not demonstrate a precision competitive to other air-shower detection techniques. This now has changed, and we will give upper limits for the precision in the next section. In addition, we have demonstrated that a reconstruction of $X_\mathrm{max}$ is possible \cite{2012ApelLOPES_MTD}, and current analyses aim at improving the reconstruction methods.

\vspace{\itemsep} \noindent \textbf{Understanding of the radio signal:}\\
To maximize the benefit of radio measurements for air-shower and cosmic-ray physics, a sufficient understanding of the radio emission process is required. LOPES \cite{FalckeNature2005} and CODALEMA \cite{Ardouin2009} confirmed that the dominant emission mechanism is the geomagnetic deflection of air-shower particles. Currently, we try to further improve the understanding of the radio emission in two ways: First, we analyze measured properties of the radio signal, like the polarization \cite{HuberARENA2012}, the lateral distribution \cite{2010ApelLOPESlateral}, and the wavefront shape \cite{SchroederIcrc2011}. Second, we compare measurements to Monte Carlo simulations of the radio emission. This way we found strong indications that on top of the geomagnetic deflection and the sub-dominant Askaryan effect \cite{Askaryan1962} (= net charge variation), also the refractive index of the air has an important influence on the radio signal, since it changes the coherence conditions for the emission \cite{deVries2011, LudwigARENA2012}.

\vspace{\itemsep} \noindent \textbf{Prototyping for large scale radio arrays:}\\
Aside from the physics goals, LOPES aims at technical developments for future, large scale radio arrays. Different antenna types \cite{KroemerSALLAIcrc2009}, self-trigger algorithms, analog and digital radio electronics have been developed and tested within the LOPES\textsuperscript{STAR} extension \cite{AschThesis2009}. The results have influenced current state-of-the-art experiments like AERA \cite{Melissas_AERA_ARENA2012} and Tunka-Rex \cite{Schroeder_TunkaRex_ARENA2012}, and now these efforts are stopped at LOPES and continued in these experiments.

\vspace{\itemsep}
In this paper we focus on the first goal, i.e.~the reconstruction of air-shower parameters. The progress in other topics is covered by LOPES proceedings in this issue \cite{HuberARENA2012, LudwigARENA2012}, or was already published in journals (see \url{www.lopes-project.org} for an overview on LOPES publications).

\begin{figure}
\centering
\includegraphics[width=0.8\columnwidth]{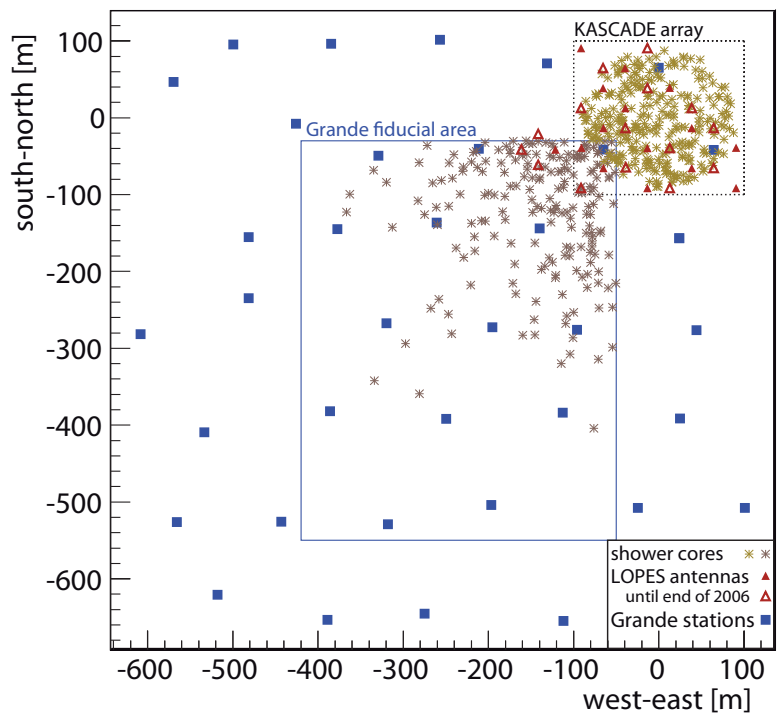}
\caption{Shower cores of events measured with the east-west aligned LOPES antennas within the fiducial areas of the KASCADE particle-detector array respectively its high energy extension 'Grande'.} \label{fig_eventMap}
\end{figure}

\section{Results}
Since the first studies on the LOPES precision for the arrival direction \cite{NiglDirection2008} and the energy \cite{HornefferIcrc2007} we made significant progress by: a) improving our amplitude and time calibration \cite{NehlsHakenjosArts2007, SchroederTimeCalibration2010}, b) introducing systematic, quantifiable quality cuts for the event selection \cite{SchroederThesis2011}, c) improving many details in our analysis software, e.g., up-sampling, and a better fit algorithm for the direction reconstruction. Due to the considerably higher statistics we used the signal of the east-west aligned antennas, only, which is analyzed under the approximation that the radio signal is purely east-west polarized. We are planning to study whether we can improve the reconstruction precision by including measurements of the differently aligned antennas. Nevertheless, already with the present analysis we can give convincing upper limits for the direction and energy precision, and are able to reconstruct the atmospheric depth of the shower maximum, $X_\mathrm{max}$. 

The reconstruction of shower parameters is based either directly on cross-correlation (CC) beamforming, or on the amplitude and arrival time in each individual antenna - after identifying the pulse with the CC-beam. During CC-beamforming the traces recorded by different antennas are shifted in time and combined to one single quantity by calculating a cross-correlation. To calculate the expected arrival times in each antenna for a given direction we assume either a spherical or a conical wavefront. The CC-beam is maximum when it is formed into the direction of the radio signal, i.e. when the time shift corresponds to the arrival time difference of the radio signal in the individual antennas (for more details see references \cite{HornefferThesis2006, HuegeARENA_LOPESSummary2010}).

In most cases the measurement precision is limited by noise \cite{SchroederNoise2010}, and only at high signal-to-noise ratios by the amplitude and time calibration of LOPES. For high signal-to-noise ratios, the uncertainty of individual amplitudes is $5\,\%$ when comparing different antennas, and another $5\,\%$ when comparing different events with each other, due to the effect of changing environmental conditions. In addition there is a scale uncertainty on the absolute LOPES amplitude of about $35\,\%$, which is relevant when comparing LOPES measurements to simulations or other experiments. The relative timing accuracy is in the order of $1\,$ns and in almost all cases negligible against the noise induced uncertainty on the pulse time of up-to $18\,$ns \cite{SchroederNoise2010}.

For the following analyses we selected events whose energy was reconstructed above $100\,$PeV by KASCADE, respectively KASCADE-Grande. Furthermore, we require a clear radio signal in the CC-beam, and several standard quality cuts \cite{SchroederThesis2011}: e.g., the core must be inside of the fiducial areas of KASCADE (281 events), respectively KASCADE-Grande (204 events). Figure \ref{fig_eventMap} displays a map of the cores of the selected LOPES events.

\subsection{Direction}

\begin{figure}
\centering
\includegraphics[width=0.7\columnwidth]{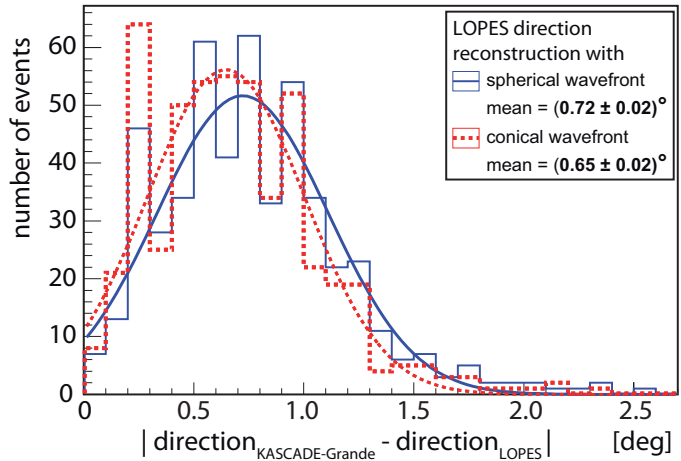}
\caption{Deviation of the KASCADE, respectively the KASCADE-Grande direction and the LOPES direction reconstructed with a spherical and a conical wavefront assumption. The mean and its statistical uncertainty has been determined by fitting a Gaussian.} \label{fig_angularResolution}
\end{figure}

For the direction reconstruction we used the KASCADE(-Grande) direction as input, and then searched within a $2.5^\circ$ neighborhood for the maximum of the CC-beam. We made several cross-checks that the $2.5^\circ$ neighborhood is sufficiently large to find the global maximum of the CC-beam: First, $2.5^\circ$ is large against the KASCADE-Grande angular resolution ($<< 1^\circ$). Second, we tried larger neighborhoods and, third, randomly varied the input direction by a value about three times larger than the LOPES angular resolution, but the results did not change significantly.

The total angular difference between the reconstructed LOPES direction and the KASCADE(-Grande) input direction is given in figure \ref{fig_angularResolution}. We take the mean difference as upper limit for the LOPES direction precision for cosmic-ray air showers. It is $0.72 \pm 0.02^\circ$ for the spherical wavefront assumption and $0.65 \pm 0.02^\circ$ for the conical wavefront assumption. In addition to the analysis we presented in reference \cite{SchroederIcrc2011}, this is a further indication that the 'true' radio wavefront is better approximated by a cone than by a sphere.

\begin{figure}[t]
\centering
\includegraphics[width=0.48\columnwidth]{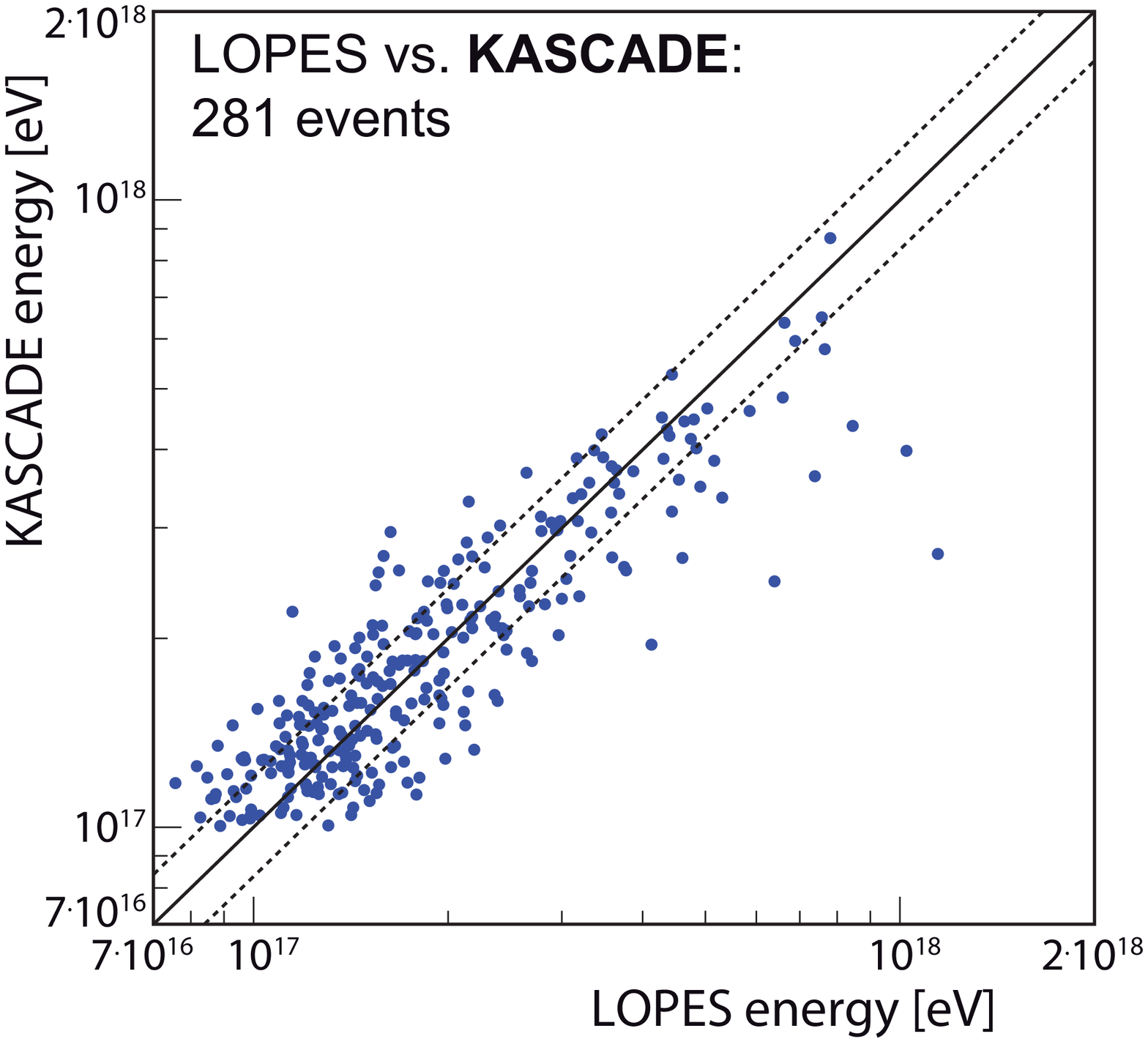}
  \hskip 0.2 cm
\includegraphics[width=0.48\columnwidth]{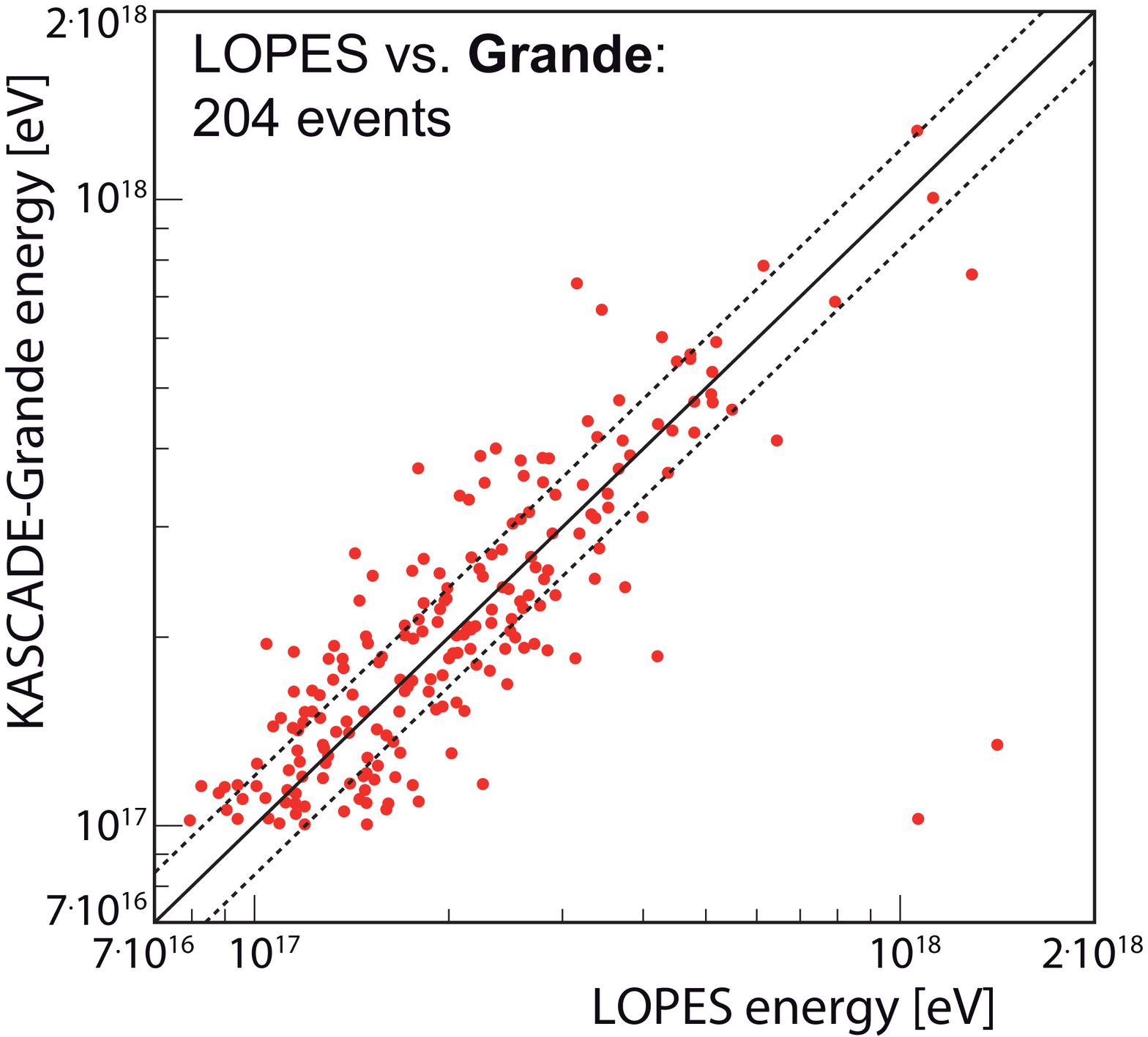}
\caption{Comparision of the LOPES reconstructed energy with the KASCADE, respectively the KASCADE-Grande energy. The dashed lines indicate an energy shift by $\pm 20\,\%$ which roughly is the KASCADE(-Grande) energy precision.} \label{fig_energyComparison}
\end{figure}

\subsection{Energy}
The energy of the primary particle can be reconstructed from the amplitude of the radio signal in several ways. By comparing the LOPES energy reconstruction with the KASCADE-(Grande) result we found that at least the statistical precision is similar or better than the KASCADE-(Grande) precision. However, we cannot determine the LOPES precision accurately, since the KASCADE-(Grande) precision is not known accurately for individual events, and because we have a selection bias. Therefore, we determine only upper limits on the LOPES energy precision. 

For the LOPES energy reconstruction we use different methods which all are essentially modifications of the formula presented by Allan \cite{Allan1971}:
\begin{equation}
E_\mathrm{Allan} \propto \frac{|\vec{\epsilon}|}{\sin{\alpha} \cdot \cos{\theta} \cdot \exp{(- d_\mathrm{axis}/d_0)}}
\end{equation}
with $E$ the energy, $\epsilon$ the radio amplitude, $\alpha$ the geomagnetic angle (= angle between the shower axis and the Earth's magnetic field), $\theta$ the zenith angle, $d_\mathrm{axis}$ the distance to the shower axis, and $d_0$ a slope parameter for an exponential lateral distribution.

One method relies on the amplitude measurements in the individual antennas, and determines the amplitude at a typical distance by fitting the lateral distribution \cite{HuegeUlrichEngel2008} (see reference \cite{PalmieriARENA2012} for new LOPES results based on this method). Another method already used for earlier LOPES results \cite{HornefferIcrc2007} is based on the amplitude of the CC-beam, i.e.~one single quantity per event combining the measurements of all antennas. We present and update here (figure \ref{fig_energyComparison}), using the following formula for energy reconstruction:
\begin{equation}
E_\mathrm{LOPES, CC} = const \cdot \frac{\epsilon_\mathrm{CC, EW} / 31\,\mathrm{MHz}}{|\vec{v} \times \vec{B}|_\mathrm{EW} \cdot \exp{(- d_\mathrm{axis}/180\,\mathrm{m})}}
\end{equation}
with $\epsilon_\mathrm{CC}$ the amplitude of the cross-correlation beam which we divide by the effective bandwidth of LOPES ($43-74\,$MHz), and $|\vec{v} \times \vec{B}|_\mathrm{EW \le 1}$ the east-west component of the geomagnetic Lorentz force unity vector, which is different from $\sin{\alpha}$ in most cases. Unlike Allan, we did not divide by $\cos \theta$, since this would increase the spread in figure \ref{fig_energyComparison}, and thus would make the energy reconstruction worse. The proportionality constant has been determined by a cross-calibration with KASCADE(-Grande) such that the mean deviation between the LOPES and KASCADE(-Grande) energy is 0. The results are $const_\mathrm{KASACDE} = 17\,$PeV$\cdot$m$\cdot$MHz$/$\textmu V, and $const_\mathrm{Grande} = 13\,$PeV$\cdot$m$\cdot$MHz$/$\textmu V. There are two possible explanations for the difference between both constants: First, in contrast to the Grande energy reconstruction, the KASCADE reconstruction has not been designed for this energy range. This might also explain the non-linearity at high energies in figure \ref{fig_energyComparison} (left panel). Second, the Grande and KASCADE events are measured at different axis distances (cf. figure \ref{fig_eventMap}), and the correction by an exponential lateral distribution with a fixed scale factor might be oversimplified to cover the full distance range.

\begin{figure}
\centering
\includegraphics[width=0.7\columnwidth]{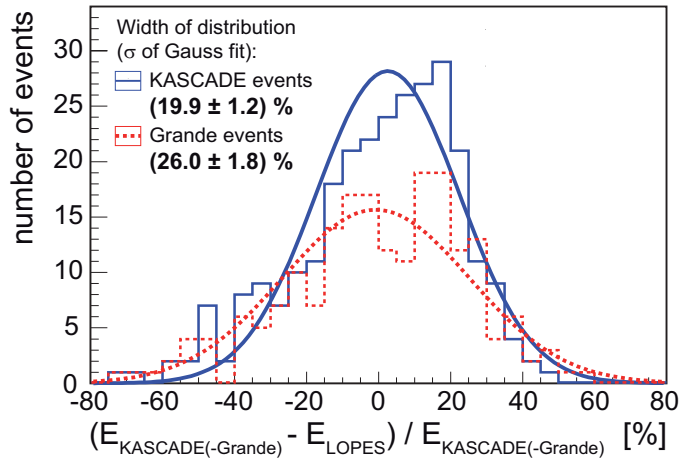}
\caption{Relative deviation of the LOPES reconstructed energy and the KASCADE, respectively the KASCADE-Grande energy. The width and its statistical uncertainty has been determined by fitting a Gaussian.} \label{fig_EnergyPrecision}
\end{figure}

Nevertheless, the used energy reconstruction formula is sufficiently good to state an upper limit for the LOPES energy precision of $20\,\%$ (figure \ref{fig_EnergyPrecision}), at least for the KASCADE events, i.e., events whose core typically is inside the LOPES array. Of course the total uncertainty is worse, since systematic uncertainties on the energy scale have to be included (see Ref.~\cite{Apel_KGenergySpectrum2012} for a discussion on the systematic uncertainties of the KASCADE-Grande energy reconstruction we used for cross-calibration). In addition, the reason for single outliers like the two events in figure \ref{fig_energyComparison} (right panel) has to be understood: Their geomagnetic angle is not exceptionally small, but they are measured at a relatively large distance from the LOPES array of several $100\,$m, and the used energy reconstruction might not be reliable in this case.

\subsection{Shower Maximum}

Recently, we provided an experimental proof that radio measurements are indeed sensitive to the longitudinal shower development, as has been theoretically assumed since long. We observe a correlation between the mean pseudorapidity of high energy muons measured by the KASCADE muon tracking detector \cite{Doll2002} and the slope of radio lateral distributions measured by LOPES \cite{2012ApelLOPES_MTD}: young showers typically have a large mean muon pseudorapidity and a steep lateral slope, and old showers vice-versa. In reference \cite{PalmieriARENA2012} we describe a method how to reconstruct the atmospheric depth of the shower maximum, $X_\mathrm{max}$, via the lateral slope. Using REAS3 simulations \cite{LudwigREAS3_2010} for calibration, we achieve an upper limit for the LOPES $X_\mathrm{max}$ precision of $90\,$g/cm\textsuperscript{2}, and the simulations indicate that in principle, i.e.~in a situation with negligible noise, a precision of better than $30\,$g/cm\textsuperscript{2} should be possible. Thus, we conclude that the main reason for the large measurement uncertainty is the high ambient noise level at the LOPES site.

We have tested an independent, second method for the reconstruction of $X_\mathrm{max}$. REAS3 simulations show that the cone angle of the wavefront is proportional to $X_\mathrm{max}$ after a correction for the shower inclination. According to the simulations, also for this method a precision of better than $30\,$g/cm\textsuperscript{2} should be achievable, and also in this case the real precision of the LOPES measurements is much worse \cite{SchroederIcrc2011}. Nevertheless, a combination of both methods might improve the measurement precision such that we are able to distinguish at least heavy from light nuclei, even in the noise environment of LOPES. In principal, there is a third method we have not tried at LOPES, since also the frequency spectrum of the radio signal ought to be sensitive to the longitudinal shower development \cite{GrebeARENA2012}. However, due to the large noise-related uncertainties of LOPES we could only demonstrate on a statistical basis that the radio amplitude decreases towards high frequencies \cite{NiglFrequencySpectrum2008, SchroederThesis2011}, but we are not able to determine the spectral slope for individual antennas on a per-event basis.

\begin{table}
\centering
\caption{LOPES reconstruction precision achieved with present analysis techniques (not including additional systematic scale uncertainties).}
\label{tab_showerCoordinates}
\begin{tabular}{lc} 
\hline
Reconstruction quantity / method & LOPES precision\\
\hline
\textbf{Direction:}&\\
Spherical CC-beamforming& $\le 0.72^\circ$\\
Conical CC-beamforming& $\le 0.65^\circ$\\
\hline
\textbf{Energy:}&\\
Amplitude of CC-beam& $\le 20\,\%$\\
Amplitude at typical axis distance& $\le 20\,\%$\\
\hline
\textbf{$X_\mathrm{max}$:}&\\
Slope of lateral distribution& $\approx 90\,$g/cm\textsuperscript{2}\\
Cone angle of radio wavefront& $\approx 200\,$g/cm\textsuperscript{2}\\
\hline
\end{tabular}
\end{table}

\section{Conclusion}
LOPES has been contributing to the development of the radio measurement technique for air showers for almost one decade, and still makes significant progress. We have a total data set of more than 500 high quality radio events which are used in many ways. Comparisons between new simulations and LOPES data reflect the improved understanding of the radio emission \cite{LudwigARENA2012}, and comparisons between the LOPES and KASCADE-Grande reconstruction of shower parameters reflect the potential of the radio method in general. For the shower direction and the energy of the primary particle, LOPES demonstrates that the radio technique can compete with the precision of established techniques. For $X_\mathrm{max}$, we developed promising methods, but the actual precision is limited by the high level of human-made background at the LOPES site. Nevertheless, experiments in regions with lower background like AERA \cite{Melissas_AERA_ARENA2012}, LOFAR \cite{NellesARENA_LOFAR2012} or Tunka-Rex \cite{Schroeder_TunkaRex_ARENA2012} should be able to achieve a precision competitive to air-fluorescence and air-Cherenkov measurements. These experiments can also test theoretical predictions that with radio arrays an energy precision below $10\,\%$ should be achievable \cite{HuegeUlrichEngel2008, PalmieriARENA2012}.

Despite all this enthusiasm, our LOPES results indicate also that the stand-alone operation of radio arrays is difficult -- at least in radio-loud environments. Self-triggering on the radio signal turned out more complicated than previously thought, since air-shower pulses are hard to distinguish from anthropogenic interferences. In addition, the absolute scale of the radio amplitude is still only understood up-to a factor of 2, which would affect both the energy scale uncertainty of stand-alone radio arrays as well as simulations of their efficiency. Nevertheless, digital radio antennas seem to be a promising extension for particle detector arrays, since radio arrays have a duty-cycle of almost $100\,\%$, and most likely can increase the precision of particle detector arrays for the energy and composition of ultra-high energy cosmic rays.

\section*{Acknowledgments}
\small{LOPES and KASCADE-Grande have been supported by the German Federal Ministry of Education and Research. KASCADE-Grande is partly supported by the MIUR and INAF of Italy, the Polish Ministry of Science and Higher Education and by the Romanian Authority for Scientific Research UEFISCDI (PNII-IDEI grant 271/2011). This research has been supported by grant number VH-NG-413 of the Helmholtz Association. We thank Johanna Lapp for improving the LOPES direction reconstruction during her Bachelor work.}

\bibliographystyle{unsrt}
\bibliography{arena2012}

\let\thefootnote\relax\footnotetext{\textsuperscript{\footnotesize{*}} Corresponding author\\Email address: frank.schroeder@kit.edu}

\end{document}